\documentclass[10pt, conference, letterpaper]{IEEEtran-v17}

\pdfoutput=1

\newlength{\figwidth}
\usepackage{xspace}
\usepackage{url}
\usepackage[cmex10]{amsmath}
\usepackage{bbm}
\usepackage{graphicx}
\usepackage{epstopdf}
\usepackage{paralist}
\usepackage[normalem]{ulem}
\usepackage{fancyref} 
\usepackage{amsmath}
\usepackage{amssymb}
\usepackage[stretch=16,shrink=16,step=4]{microtype}
\usepackage{color}
\definecolor{links}{rgb}{0.7,0,0}   
\definecolor{urls}{rgb}{0,0,0.8}    
\definecolor{cites}{rgb}{0,0,0.8}   
\usepackage{dsfont}
\usepackage{footnote}
\usepackage{balance}
\usepackage{float}
\usepackage{cite}
\usepackage{wasysym}
\usepackage{amssymb}
\let\labelindent\relax
\usepackage[inline]{enumitem}
\usepackage{setspace}
\usepackage{tabularx,booktabs}
\newcolumntype{L}[1]{>{\raggedright\let\newline\\\arraybackslash\hspace{0pt}}m{#1}}
\newcolumntype{C}[1]{>{\centering\let\newline\\\arraybackslash\hspace{0pt}}m{#1}}
\newcolumntype{R}[1]{>{\raggedleft\let\newline\\\arraybackslash\hspace{0pt}}m{#1}}
\usepackage{boldline}

\usepackage{vmr-symbols-vecbold}
\usepackage{standard-macros}
\usepackage{mathbbol}
\usepackage{hyperref}

\usepackage{tikz}
\usetikzlibrary{patterns}
\usetikzlibrary{shapes,arrows,positioning}
\usetikzlibrary{calc}
\usetikzlibrary{fit}
\usetikzlibrary{plotmarks}
\tikzset{every picture/.style={font issue=\scriptsize, >=stealth},font issue/.style={execute at begin picture={#1\selectfont}}}
\tikzset{three sided left/.style={
        draw=none,
        xshift=\pgflinewidth,
        append after command={
            [shorten <= -0.5\pgflinewidth]
            ([shift={(-1.5\pgflinewidth,-0.5\pgflinewidth)}]\tikzlastnode.north east) edge ([shift={( 0.5\pgflinewidth,-0.5\pgflinewidth)}]\tikzlastnode.north west) 
            ([shift={( 0.5\pgflinewidth,-0.5\pgflinewidth)}]\tikzlastnode.north west) edge ([shift={( 0.5\pgflinewidth,+0.5\pgflinewidth)}]\tikzlastnode.south west)            
            ([shift={( 0.5\pgflinewidth,+0.5\pgflinewidth)}]\tikzlastnode.south west) edge ([shift={(-1.0\pgflinewidth,+0.5\pgflinewidth)}]\tikzlastnode.south east)
        }}}
        
\tikzset{three sided right/.style={
        draw=none,
        xshift=-\pgflinewidth,
        append after command={
            [shorten <= -0.5\pgflinewidth]
            ([shift={( 1.5\pgflinewidth,-0.5\pgflinewidth)}]\tikzlastnode.north west) edge ([shift={(-0.5\pgflinewidth,-0.5\pgflinewidth)}]\tikzlastnode.north east) 
            ([shift={(-0.5\pgflinewidth,-0.5\pgflinewidth)}]\tikzlastnode.north east) edge ([shift={(-0.5\pgflinewidth,+0.5\pgflinewidth)}]\tikzlastnode.south east)            
            ([shift={(-0.5\pgflinewidth,+0.5\pgflinewidth)}]\tikzlastnode.south east) edge ([shift={( 1.0\pgflinewidth,+0.5\pgflinewidth)}]\tikzlastnode.south west)
        }}}

\usepackage{pgfplots}
\usepgfplotslibrary{fillbetween}
\pgfplotsset{
  compat=newest, 
  width=\columnwidth,    
  height=0.8\columnwidth,   
  plot coordinates/math parser=false,
  standard/.style={
    axis equal,
    axis line style=help lines,
    axis x line=center,
    axis y line=center,
    axis z line=center},
    grid style={dashed,gray},
    minor grid style={dotted,gray},
    major grid style={dotted,gray},
    ylabel absolute, ylabel style={yshift=-0.4cm},
    xlabel absolute, xlabel style={yshift=0.25cm}
}

\makeatletter
 \pgfdeclarepatternformonly[\tikz@pattern@color,\LineSpace,\LineWidth]{my horizontal lines}%
    {\pgfpointorigin}{\pgfqpoint{100pt}{1pt}}{\pgfqpoint{100pt}{\LineSpace}}%
    {
        \pgfsetcolor{\tikz@pattern@color}
        \pgfsetlinewidth{\LineWidth}
        \pgfpathmoveto{\pgfqpoint{0pt}{0.5pt}}
        \pgfpathlineto{\pgfqpoint{100pt}{0.5pt}}
        \pgfusepath{stroke}
    }
 
 \pgfdeclarepatternformonly[\tikz@pattern@color,\LineSpace,\LineWidth]{my vertical lines}%
    {\pgfpointorigin}{\pgfqpoint{1pt}{100pt}}{\pgfqpoint{\LineSpace}{100pt}}%
    {
        \pgfsetcolor{\tikz@pattern@color}
        \pgfsetlinewidth{\LineWidth}
        \pgfpathmoveto{\pgfqpoint{0.5pt}{0pt}}
        \pgfpathlineto{\pgfqpoint{0.5pt}{100pt}}
        \pgfusepath{stroke}
    } 
 
 \pgfdeclarepatternformonly[\tikz@pattern@color,\LineSpace,\LineWidth]{my grid}%
    {\pgfqpoint{-1pt}{-1pt}}{\pgfqpoint{\LineSpace}{\LineSpace}}
    {\pgfqpoint{\LineSpace}{\LineSpace}}%
    {
        \pgfsetcolor{\tikz@pattern@color}
        \pgfsetlinewidth{\LineWidth}
        \pgfpathmoveto{\pgfqpoint{0pt}{0pt}}
        \pgfpathlineto{\pgfqpoint{0pt}{\LineSpace + 0.1pt}}
        \pgfpathmoveto{\pgfqpoint{0pt}{0pt}}
        \pgfpathlineto{\pgfqpoint{\LineSpace + 0.1pt}{0pt}}
        \pgfusepath{stroke}
    }
 
 \pgfdeclarepatternformonly[\tikz@pattern@color,\LineSpace,\LineWidth]{my north east lines}
    {\pgfqpoint{-\LineWidth}{-\LineWidth}}{\pgfqpoint{\LineSpace}{\LineSpace}}
    {\pgfqpoint{\LineSpace}{\LineSpace}}%
    {
        \pgfsetcolor{\tikz@pattern@color}
        \pgfsetlinewidth{\LineWidth}
        \pgfpathmoveto{\pgfqpoint{-\LineWidth}{-\LineWidth}}
        \pgfpathlineto{\pgfqpoint{\LineSpace + 0.1pt}{\LineSpace + 0.1pt}}
        \pgfusepath{stroke}
    }
 
 \pgfdeclarepatternformonly[\tikz@pattern@color,\LineSpace,\LineWidth]{my north west lines}
    {\pgfqpoint{-\LineWidth}{-\LineWidth}}{\pgfqpoint{\LineSpace}{\LineSpace}}
    {\pgfqpoint{\LineSpace}{\LineSpace}}%
    {
        \pgfsetcolor{\tikz@pattern@color}
        \pgfsetlinewidth{\LineWidth}
        \pgfpathmoveto{\pgfqpoint{-\LineWidth}{\LineSpace}}
        \pgfpathlineto{\pgfqpoint{\LineSpace + 0.1pt}{-\LineWidth}}
        \pgfusepath{stroke}
    }
 
 \pgfdeclarepatternformonly[\tikz@pattern@color,\LineSpace,\LineWidth]{my crosshatch}%
    {\pgfqpoint{-1pt}{-1pt}}{\pgfqpoint{\LineSpace}{\LineSpace}}
    {\pgfqpoint{\LineSpace}{\LineSpace}}%
    {
        \pgfsetcolor{\tikz@pattern@color}
        \pgfsetlinewidth{\LineWidth}
        \pgfpathmoveto{\pgfqpoint{\LineSpace + 0.1pt}{0pt}}
        \pgfpathlineto{\pgfqpoint{0pt}{\LineSpace + 0.1pt}}
        \pgfpathmoveto{\pgfqpoint{0pt}{0pt}}
        \pgfpathlineto{\pgfqpoint{\LineSpace + 0.1pt}{\LineSpace + 0.1pt}}
        \pgfusepath{stroke}
    }
 
 \pgfdeclarepatternformonly[\tikz@pattern@color,\LineSpace,\PointSize]{my dots}%
    {\pgfqpoint{-\LineSpace*0.25}{-\LineSpace*0.25}}
    {\pgfqpoint{\LineSpace*0.25}{\LineSpace*0.25}}
    {\pgfqpoint{\LineSpace*0.75}{\LineSpace*0.75}}%
    {
        \pgfsetcolor{\tikz@pattern@color}
        \pgfpathcircle{\pgfqpoint{0pt}{0pt}}{\PointSize}
        \pgfusepath{fill}
    }
\makeatother
 
\newdimen\LineSpace
\newdimen\PointSize
\newdimen\LineWidth
\tikzset{
    line space/.code={\LineSpace=#1},
    line space=3pt
}
\tikzset{
    point size/.code={\PointSize=#1},
    point size=.5pt
}
\tikzset{
    pattern line width/.code={\LineWidth=#1},
    pattern line width=.4pt
}

\DeclareSymbolFontAlphabet{\amsmathbb}{AMSb}%

\newcommand{\lro}[1]{\lefto({#1}\right)}																

\newcommand{\lr}[1]{\left({#1}\right)}																
\newcommand{\lrb}[1]{\left \lbrace {#1} \right \rbrace}															

\safemath{\dopplerspread}{B_D}																								
\safemath{\delayspread}{T_D}																									
\safemath{\nc}{n\sub{c}}																										
\safemath{\nf}{n\sub{f}}																										
\safemath{\efa}{p\sub{sc}}
\safemath{\efb}{p\sub{cs}}
\safemath{\ef}{\epsilon\sub{f}	}
\safemath{\nd}{n\sub{d}}																										
\safemath{\ntx}{n\sub{t}} 																											
\safemath{\nrx}{n\sub{r}}																											
\safemath{\ntxt}{\tilde{n\sub{t}}}																											
\safemath{\cb}{\ensuremath{L}} 																								
\safemath{\cl}{\ensuremath{n}} 																								
\safemath{\txanto}{{\ensuremath{\tilde{m}_t}}} 																		
\safemath{\cs}{M} 																														
\safemath{\idPustm}{\ensuremath{S_{k}}}
\safemath{\error}{\ensuremath{\epsilon}} 																				
\safemath{\eexp}{\ensuremath{\mathcal{E}}} 																			
\safemath{\nsubc}{n\sub{s}}			 																						
\safemath{\nofdm}{n\sub{o}} 																									
\safemath{\bc}{\ensuremath{B_c}} 																							
\safemath{\ts}{\ensuremath{T_s}} 																							
\safemath{\nrb}{\ensuremath{n_{rb}}} 																						
\safemath{\rul}{\ensuremath{\rho\sub{ul}}}
\safemath{\rdl}{\ensuremath{\rho\sub{dl}}}

\safemath{\nres}{\ell}
\safemath{\nr}{n\sub{r}}
\newcommand{\cgauss}[2]{\mathcal{CN}\lro{\ensuremath{#1, #2}  }}   								
\safemath{\maxk}{M^*\lr{\nres, \nsubc, \nofdm, \epsilon, \rho}}
\safemath{\Rmax}{R^*}
\safemath{\Emin}{E\sub{b}^*/N_0}
\safemath{\Eminf}{\frac{E\sub{b}^*}{N_0}}
\safemath{\np}{\ensuremath{n\sub{p}}}
\safemath{\ndf}{\ensuremath{\bar{n}\sub{d}}}
\safemath{\npf}{\ensuremath{\bar{n}\sub{p}}}
\safemath{\code}{\ensuremath{\mathcal{C}}}
\safemath{\err}{\ensuremath{\epsilon}}
\safemath{\rp}{\ensuremath{\rho\sub{p}}}
\safemath{\rd}{\ensuremath{\rho\sub{d}}}
\safemath{\cohtime}{\ensuremath{T\sub{c}}}
\safemath{\cohbw}{\ensuremath{B\sub{c}}}
\safemath{\nmax}{\ensuremath{\ell\sub{m}}}
\safemath{\ntot}{\ensuremath{n\sub{tot}}}

\safemath{\yp}{\ensuremath{\randvecy_{\nu}^{(\text{p})}}}
\safemath{\yd}{\ensuremath{\randvecy_{\nu}^{(\text{d})}}}
\safemath{\ypd}{\ensuremath{\vecy_{\nu}^{(\text{p})}}}
\safemath{\ydd}{\ensuremath{\vecy_{\nu}^{(\text{d})}}}

\safemath{\ypf}{\ensuremath{\bar{\randvecy}_{\nu}^{(\text{p})}}}
\safemath{\ydf}{\ensuremath{\bar{\randvecy}_{\nu}^{(\text{d})}}}
\safemath{\ypdf}{\ensuremath{\bar{\vecy}_{\nu}^{(\text{p})}}}
\safemath{\yddf}{\ensuremath{\bar{\vecy}_{\nu}^{(\text{d})}}}

\safemath{\xp}{\ensuremath{\vecx^{(\text{p})}}}
\safemath{\xd}{\ensuremath{\randvecx_{\nu}^{(\text{d})}}}
\safemath{\xdd}{\ensuremath{\vecx_{\nu}^{(\text{d})}}}

\safemath{\xpf}{\ensuremath{\bar{\vecx}^{(\text{p})}}}
\safemath{\xdf}{\ensuremath{\bar{\randvecx}_{\nu}^{(\text{d})}}}
\safemath{\xddf}{\ensuremath{\bar{\vecx}_{\nu}^{(\text{d})}}}

\safemath{\xdb}{\ensuremath{\overline{\randvecx}^{(\text{d})}}}
\safemath{\Pxd}{\ensuremath{P_{\randvecx^{(\text{d})}}}}

\safemath{\xpbar}{\ensuremath{\overline{\matX}^{(\text{p})}}}
\safemath{\xdbar}{\ensuremath{\overline{\randmatX}^{(\text{d})}}}

\safemath{\xdv}{\ensuremath{\randvecx^{(\text{d})}}}
\safemath{\xdbarv}{\ensuremath{\overline{\randvecx}^{(\text{d})}}}
\safemath{\ydv}{\ensuremath{\randvecy^{(\text{d})}}}

\safemath{\xdr}{\ensuremath{\matX^{(\text{d})}}}

\safemath{\ttx}{\ensuremath{\tau\sub{tx}}}
\safemath{\trx}{\ensuremath{\tau\sub{rx}}}
\safemath{\ack}{\ensuremath{\mathrm{s}}}
\safemath{\nack}{\ensuremath{\mathrm{c}}}

\safemath{\mI}{\ensuremath{i\lro{\randvecy ; \randvecx}}} 				

\safemath{\randveca}{\bm{A}}
\safemath{\randvecb}{\bm{B}}
\safemath{\randvecc}{\bm{C}}
\safemath{\randvecd}{\bm{D}}
\safemath{\randvece}{\bm{E}}
\safemath{\randvecf}{\bm{F}}
\safemath{\randvecg}{\bm{G}}
\safemath{\randvech}{\bm{H}}
\safemath{\randveci}{\bm{I}}
\safemath{\randvecj}{\bm{J}}
\safemath{\randveck}{\bm{K}}
\safemath{\randvecl}{\bm{L}}
\safemath{\randvecm}{\bm{M}}
\safemath{\randvecn}{\bm{N}}
\safemath{\randveco}{\bm{O}}
\safemath{\randvecp}{\bm{P}}
\safemath{\randvecq}{\bm{Q}}
\safemath{\randvecr}{\bm{R}}
\safemath{\randvecs}{\bm{S}}
\safemath{\randvect}{\bm{T}}
\safemath{\randvecu}{\bm{U}}
\safemath{\randvecv}{\bm{V}}
\safemath{\randvecw}{\bm{W}}
\safemath{\randvecx}{\bm{X}}
\safemath{\randvecy}{\bm{Y}}
\safemath{\randvecz}{\bm{Z}}
\safemath{\randvecphi}{\bm{\Phi}}

\safemath{\randmatA}{\amsmathbb{A}}
\safemath{\randmatB}{\amsmathbb{B}}
\safemath{\randmatC}{\amsmathbb{C}}
\safemath{\randmatD}{\amsmathbb{D}}
\safemath{\randmatE}{\amsmathbb{E}}
\safemath{\randmatF}{\amsmathbb{F}}
\safemath{\randmatG}{\amsmathbb{G}}
\safemath{\randmatH}{\amsmathbb{H}}
\safemath{\randmatI}{\amsmathbb{I}}
\safemath{\randmatJ}{\amsmathbb{J}}
\safemath{\randmatK}{\amsmathbb{K}}
\safemath{\randmatL}{\amsmathbb{L}}
\safemath{\randmatM}{\amsmathbb{M}}
\safemath{\randmatN}{\amsmathbb{N}}
\safemath{\randmatO}{\amsmathbb{O}}
\safemath{\randmatP}{\amsmathbb{P}}
\safemath{\randmatQ}{\amsmathbb{Q}}
\safemath{\randmatR}{\amsmathbb{R}}
\safemath{\randmatS}{\amsmathbb{S}}
\safemath{\randmatT}{\amsmathbb{T}}
\safemath{\randmatU}{\amsmathbb{U}}
\safemath{\randmatV}{\amsmathbb{V}}
\safemath{\randmatW}{\amsmathbb{W}}
\safemath{\randmatX}{\amsmathbb{X}}
\safemath{\randmatY}{\amsmathbb{Y}}
\safemath{\randmatZ}{\amsmathbb{Z}}
\safemath{\randmatSigma}{\mathbb{\Sigma}}
\safemath{\randmatPhi}{\mathbb{\Phi}}
\safemath{\randmatLambda}{\mathbb{\Lambda}}

\safemath{\matSigma}{\bm{\Sigma}}
\safemath{\matPhi}{\bm{\Phi}}
\safemath{\matLambda}{\bm{\Lambda}}

\usepackage{glossaries}
\glsdisablehyper
\makeglossaries
\newglossaryentry{mmse}
{
  name={MMSE},
  description={minimum mean-square error},
  first={\glsentrydesc{mmse} (\glsentrytext{mmse})}
}

\newglossaryentry{mrc}
{
  name={MRC},
  description={maximum ratio combining},
  first={\glsentrydesc{mrc} (\glsentrytext{mrc})}
}

\newglossaryentry{ue}
{
  name={UE},
  description={user equipment},
  first={\glsentrydesc{ue} (\glsentrytext{ue})},
  plural={UEs},
  descriptionplural={user equipments},
  firstplural={\glsentrydescplural{ue} (\glsentryplural{ue})}
}

\newglossaryentry{bs}
{
  name={BS},
  description={base station},
  first={\glsentrydesc{bs} (\glsentrytext{bs})},
  plural={BS},
  descriptionplural={base stations},
  firstplural={\glsentrydescplural{bs} (\glsentryplural{bs})}
}

\newglossaryentry{ostbc}
{
  name={OSTBC},
  description={orthogonal space-time block code},
  first={\glsentrydesc{ostbc} (\glsentrytext{ostbc})},
  plural={OSTBCs},
  descriptionplural={orthogonal space-time block codes},
  firstplural={\glsentrydescplural{ostbc} (\glsentryplural{ostbc})}
}

\newglossaryentry{biawgn}
{
  name={bi-AWGN},
  description={binary-input additive white Gaussian noise},
  first={\glsentrydesc{biawgn} (\glsentrytext{biawgn})}
}

\newglossaryentry{flnf}
{
  name={FLNF},
  description={fixed-length no-feedback},
  first={\glsentrydesc{flnf} (\glsentrytext{flnf})}
}

\newglossaryentry{crc}
{
  name={CRC},
  description={cyclic redundancy check},
  first={\glsentrydesc{crc} (\glsentrytext{crc})}
}

\newglossaryentry{bsc}
{
  name={BSC},
  description={binary symmetric channel},
  first={\glsentrydesc{bsc} (\glsentrytext{bsc})}
  plural={BSCs},
  descriptionplural={binary symmetric channels},
  firstplural={\glsentrydescplural{bsc} (\glsentryplural{bsc})}
}

\newglossaryentry{bec}
{
  name={BEC},
  description={binary-erasure channel},
  first={\glsentrydesc{bec} (\glsentrytext{bec})}
}
\newglossaryentry{awgn}
{
  name={AWGN},
  description={additive white Gaussian noise},
  first={\glsentrydesc{awgn} (\glsentrytext{awgn})}
}
\newglossaryentry{3gpp}
{
  name={3GPP},
  description={3rd generation partnership project},
  first={\glsentrydesc{3gpp} (\glsentrytext{3gpp})}
}

\newglossaryentry{dl}
{
  name={DL},
  description={downlink},
  first={\glsentrydesc{dl} (\glsentrytext{dl})}
}

\newglossaryentry{LED}
{
  name={LED},
  description={light emitting diode},
  first={\glsentrydesc{LED} (\glsentrytext{LED})},
  plural={LEDs},
  descriptionplural={light emitting diodes},
  firstplural={\glsentrydescplural{LED} (\glsentryplural{LED})}
}

\newglossaryentry{lte}
{
  name={LTE},
  description={long term evolution},
  first={\glsentrydesc{lte} (\glsentrytext{lte})}
}
\newglossaryentry{ofdm}
{
  name={OFDM},
  description={orthogonal frequency-division multiplexing},
  first={\glsentrydesc{ofdm} (\glsentrytext{ofdm})}
}

\newglossaryentry{ul}
{
  name={UL},
  description={uplink},
  first={\glsentrydesc{ul} (\glsentrytext{ul})}
}
\newglossaryentry{rb}
{
  name={RB},
  description={resource block},
  first={\glsentrydesc{rb} (\glsentrytext{rb})},
  plural={RBs},
  descriptionplural={resource blocks},
  firstplural={\glsentrydescplural{rb} (\glsentryplural{rb})}
}
\newglossaryentry{cu}
{
  name={CU},
  description={channel use},
  first={\glsentrydesc{cu} (\glsentrytext{cu})},
  plural={CUs},
  descriptionplural={channel uses},
  firstplural={\glsentrydescplural{cu} (\glsentryplural{cu})}
}
\newglossaryentry{tti}
{
  name={TTI},
  description={transmission time interval},
  first={\glsentrydesc{tti} (\glsentrytext{tti})},
  plural={TTI's},
  descriptionplural={transmission time intervals},
  firstplural={\glsentrydescplural{tti} (\glsentryplural{tti})}
}
\newglossaryentry{iid}
{
  name={iid},
  description={independent and identically distributed},
  first={\glsentrydesc{iid} (\glsentrytext{iid})}
}
\newglossaryentry{psd}
{
  name={PSD},
  description={power spectral density},
  first={\glsentrydesc{psd} (\glsentrytext{psd})}
}

\newglossaryentry{mimo}
{
  name={MIMO},
  description={multiple-input multiple-output},
  first={\glsentrydesc{mimo} (\glsentrytext{mimo})}
}

\newglossaryentry{bler}
{
  name={BLER},
  description={block error probability},
  first={\glsentrydesc{bler} (\glsentrytext{bler})}
}
\newglossaryentry{mtc}
{
  name={MTC},
  description={machine-type communication},
  first={\glsentrydesc{mtc} (\glsentrytext{mtc})}
}
\newglossaryentry{csi}
{
  name={CSI},
  description={channel state information},
  first={\glsentrydesc{csi} (\glsentrytext{csi})}
}
\newglossaryentry{dvbt}
{
  name={DVB-T},
  description={terrestrial digital video broadcasting},
  first={\glsentrydesc{dvbt} (\glsentrytext{dvbt})}
}
\newglossaryentry{pat}
{
  name={PAT},
  description={pilot-assisted transmission},
  first={\glsentrydesc{pat} (\glsentrytext{pat})}
}
\newglossaryentry{ml}
{
  name={ML},
  description={maximum likelihood},
  first={\glsentrydesc{ml} (\glsentrytext{ml})}
}
\newglossaryentry{dt}
{
  name={DT},
  description={dependence-testing},
  first={\glsentrydesc{dt} (\glsentrytext{dt})}
}
\newglossaryentry{ustm}
{
  name={USTM},
  description={unitary space-time modulation},
  first={\glsentrydesc{ustm} (\glsentrytext{ustm})}
}
\newglossaryentry{siso}
{
  name={SISO},
  description={single-input single-output},
  first={\glsentrydesc{siso} (\glsentrytext{siso})}
}
\newglossaryentry{snn}
{
  name={SNN},
  description={scaled nearest-neighbor},
  first={\glsentrydesc{snn} (\glsentrytext{snn})}
}

\newglossaryentry{snnd}
{
  name={SNND},
  description={scaled nearest-neighbor decoding},
  first={\glsentrydesc{snnd} (\glsentrytext{snnd})}
}
\newglossaryentry{rcus}
{
  name={RCUs},
  description={random-coding union bound with parameter $s$},
  first={\glsentrydesc{rcus} (\glsentrytext{rcus})}
}
\newglossaryentry{osd}
{
  name={OSD},
  description={ordered statistics decoding},
  first={\glsentrydesc{osd} (\glsentrytext{osd})}
}
\newglossaryentry{llr}
{
  name={LLR},
  description={log-likelihood ratio},
  first={\glsentrydesc{llr} (\glsentrytext{llr})}
   plural={LLR's},
  descriptionplural={log-likelihood ratios},
  firstplural={\glsentrydescplural{llr} (\glsentryplural{llr})}
}
\newglossaryentry{qpsk}
{
  name={QPSK},
  description={quaternary phase shift keying},
  first={\glsentrydesc{qpsk} (\glsentrytext{qpsk})}
}
\newglossaryentry{nlos}
{
  name={NLOS},
  description={non-line-of-sight},
  first={\glsentrydesc{nlos} (\glsentrytext{nlos})}
}
\newglossaryentry{los}
{
  name={LOS},
  description={line-of-sight},
  first={\glsentrydesc{los} (\glsentrytext{los})}
}
\newglossaryentry{cgf}
{
  name={CGF},
  description={cumulant-generating function},
  first={\glsentrydesc{cgf} (\glsentrytext{cgf})}
   plural={CGF's},
  descriptionplural={cumulant-generating functions},
  firstplural={\glsentrydescplural{cgf} (\glsentryplural{cgf})}
}
\newglossaryentry{pdf}
{
  name={PDF},
  description={probability density function},
  first={\glsentrydesc{pdf} (\glsentrytext{pdf})}
   plural={PDF's},
  descriptionplural={probability density functions},
  firstplural={\glsentrydescplural{pdf} (\glsentryplural{pdf})}
}

\newglossaryentry{ccdf}
{
  name={CCDF},
  description={complementary cumulative distribution function},
  first={\glsentrydesc{ccdf} (\glsentrytext{ccdf})}
}

\newglossaryentry{cdf}
{
  name={CDF},
  description={cumulative distribution function},
  first={\glsentrydesc{cdf} (\glsentrytext{cdf})}
}

\newglossaryentry{gmi}
{
  name={GMI},
  description={generalized mutual information},
  first={\glsentrydesc{gmi} (\glsentrytext{gmi})}
}

\newglossaryentry{arq}
{
  name={ARQ},
  description={automatic repeat request},
  first={\glsentrydesc{arq} (\glsentrytext{arq})}
}

\newglossaryentry{harq}
{
  name={HARQ},
  description={hybrid automatic repeat request},
  first={\glsentrydesc{harq} (\glsentrytext{harq})}
}

\newglossaryentry{fbl}
{
  name={FBL-NF},
  description={fixed blocklength with no feedback},
  first={\glsentrydesc{fbl} (\glsentrytext{fbl})}
}
\newglossaryentry{ssnf}
{
  name={SS-NF},
  description={single-shot no-feedback},
  first={\glsentrydesc{ssnf} (\glsentrytext{ssnf})}
}

\newglossaryentry{urllc}
{
  name={URLLC},
  description={ultra-reliable low-latency communications},
  first={\glsentrydesc{urllc} (\glsentrytext{urllc})}
}

\newglossaryentry{vlsf}
{
  name={VLSF},
  description={Variable-length stop-feedback},
  first={\glsentrydesc{vlsf} (\glsentrytext{vlsf})}
}

\newglossaryentry{vlnsf}
{
  name={VLNSF},
  description={variable-length noisy stop feedback},
  first={\glsentrydesc{vlnsf} (\glsentrytext{vlnsf})}
}

\newglossaryentry{dmt}
{
  name={DMT},
  description={diversity-multiplexing tradeoff},
  first={\glsentrydesc{dmt} (\glsentrytext{dmt})}
}

\newglossaryentry{dmdt}
{
  name={DMDT},
  description={diversity-multiplexing-delay tradeoff},
  first={\glsentrydesc{dmdt} (\glsentrytext{dmdt})}
}

\newglossaryentry{tddt}
{
  name={TDDT},
  description={throughput-diversity-delay tradeoff},
  first={\glsentrydesc{tddt} (\glsentrytext{tddt})}
}

\newglossaryentry{tdd}
{
  name={TDD},
  description={time-division duplex},
  first={\glsentrydesc{tdd} (\glsentrytext{tdd})}
}

\newglossaryentry{stbc}
{
  name={STBC},
  description={space-time block-codes},
  first={\glsentrydesc{stbc} (\glsentrytext{stbc})},
  plural={STBC's},
  descriptionplural={space-time block-codes},
  firstplural={\glsentrydescplural{stbc} (\glsentryplural{stbc})}
}

\newglossaryentry{harqir}
{
  name={HARQ-IR},
  description={hybrid automatic repeat request with incremental redundancy},
  first={\glsentrydesc{harqir} (\glsentrytext{harqir})}
}

\newglossaryentry{harqcc}
{
  name={HARQ-CC},
  description={hybrid automatic repeat request with chase combining},
  first={\glsentrydesc{harqcc} (\glsentrytext{harqcc})}
}

\newglossaryentry{wp1}
{
  name={w.p.1.},
  description={with probability one},
  first={\glsentrydesc{wp1} (\glsentrytext{wp1})}
}

\newglossaryentry{vlff}
{
  name={VLFF},
  description={variable-length full feedback},
  first={\glsentrydesc{vlff} (\glsentrytext{vlff})}
}

\newglossaryentry{dmc}
{
  name={DMC},
  description={discrete memoryless channel},
  first={\glsentrydesc{dmc} (\glsentrytext{dmc})}
  plural={DMCs},
  descriptionplural={discrete memoryless channels},
  firstplural={\glsentrydescplural{dmc} (\glsentryplural{dmc})}
}

\usepackage{pgfplots}
\usepgfplotslibrary{groupplots}
\usetikzlibrary{pgfplots.groupplots}

\usepgfplotslibrary{external}

\newcommand{\fwidth}{\textwidth}

\makeatletter
\def\@IEEEinterspaceratioM{0.265}
\def\@IEEEinterspaceMINratioM{0.1651}
\def\@IEEEinterspaceMAXratioM{0.38}

\def\@IEEEinterspaceratioB{0.31}
\def\@IEEEinterspaceMINratioB{0.19}
\def\@IEEEinterspaceMAXratioB{0.38}
\@IEEEtunefonts
\makeatother
\let\abs\undefined
\newcommand{\abs}[1]{\lvert#1\rvert}		

\definecolor{red}{RGB}{220, 10, 10}
\definecolor{blue}{RGB}{10, 40, 200}
\definecolor{green}{RGB}{10, 200, 10}
\definecolor{orange}{RGB}{255, 165, 0}
\definecolor{pink}{RGB}{255, 51, 204}

\pgfplotscreateplotcyclelist{colorbrewer}{
{red},
{blue},
{green},
{orange},
{pink}
}
\pgfplotscreateplotcyclelist{colorfour}{
{red},
{blue},
{green},
{orange}
}

\let\abs\undefined
\newcommand{\abs}[1]{\lvert#1\rvert}		

\let\marginnote\undefined

\newcommand{\marginnote}[1]{\marginpar{\framebox{\framebox{#1}}}}

\allowdisplaybreaks
\begin{document}
\IEEEoverridecommandlockouts

\title{Short-Packet Transmission over a Bidirectional Massive MIMO link}
\author{\IEEEauthorblockN{Johan \"Ostman, Alejandro Lancho, and Giuseppe Durisi}
\IEEEauthorblockA{
 Chalmers University of Technology, Gothenburg, Sweden}
\thanks{This work was supported by the Swedish Research Council under grants 2014-6066 and 2016-03293.}
 }
 \maketitle

\begin{abstract}
We consider the transmission of short packets over a bidirectional communication link where multiple devices, e.g., sensors and actuators, exchange  small-data payloads with a base station equipped with a large antenna array.
Using results from finite-blocklength information theory, we characterize the minimum SNR required to achieve a target error probability for a fixed packet length and a fixed payload size.
Our nonasymptotic analysis, which applies to the scenario in which  the bidirectional communication is device-initiated, and also to the more challenging case when it is base-station initiated,
provides guidelines on the design of massive multiple-input multiple-output links that need to support sporadic ultra-reliable low-latency transmissions.
Specifically, it allows us to determine the optimal amount of resources that need to be dedicated to the acquisition of channel state information.
\end{abstract}

\section{Introduction}
Because of its ability to accommodate many parallel high-throughput links in the same time-frequency resources, massive \gls{mimo} has been identified as a key technology for next-generation wireless systems~\cite{boccardi14-02a,zaidi18-a}.
Furthermore, the potentially large spatial diversity provided by massive \gls{mimo} makes this technology also relevant for some of the new use cases in next generation's wireless systems, where reliability and latency, rather than throughput, are in focus~\cite{Popovski19}.

One such use case is \gls{urllc}, where small data payloads need to be transmitted under stringent latency and  reliability constraints.
For example, in the context of factory automation, one may need to deliver packets of $100$ bits, conveying, e.g., readings from sensors or commands to actuators, within hundreds of microseconds and with a reliability no smaller than $99.999\%$.
In this scenario, the stringent delay constraint prohibits the exploitation of diversity in time; furthermore, there may be only limited diversity in frequency.
Thus, spatial diversity offered by multiple antennas is critical to achieve the desired reliability~\cite{johansson15-06a}.

The purpose of this paper is to provide a characterization of the error probability achievable in a bidirectional massive MIMO link as a function of the SNR, the number of active \glspl{ue}, the number of the available antennas at the \gls{bs}, and the size of the information payload.
Previous results reported in the literature~\cite{bana18-10a,bana19-05a,Karlsson18} rely on asymptotic performance metrics, such as ergodic and outage capacity, to characterize the performance of latency-constrained communication systems.
Our analysis relies instead on tools from finite-blocklength information theory, which are more suited to the blocklengths of interest in \gls{urllc} than asymptotic performance metrics~\cite{durisi16-02a}.

\paragraph*{Literature review}
Most of the information-theoretic characterizations of  massive MIMO communication links deal with bounds on the ergodic capacity~\cite{Bjornsson17}.
These bounds are typically obtained under specific assumptions on the signaling scheme and on the operations performed at the receiver side, which are motivated by practical considerations.
Specifically, it is common to postulate that the system operates in \gls{tdd} mode, that pilot symbols are transmitted in the \gls{ul}, and that the \gls{bs} performs \gls{mmse} channel estimation followed by linear combining in the \gls{ul} and linear precoding in the \gls{dl}.
In both the \gls{ul} and the \gls{dl}, the channel estimate is treated as perfect.
Furthermore, the most commonly used bounds assume implicitly that the receiver (be it the \gls{bs} or the \gls{ue}) performs \emph{mismatched scaled-nearest-neighbor decoding}~\cite{lapidoth02-05a} by treating channel estimation errors and residual multiuser interference as noise.

These ergodic bounds are, however, unsuitable for \gls{urllc}.
Indeed, they rely on the assumption that each codeword spans a large number of diversity branches over time---an assumption that is not valid in low-latency scenarios.

An alternative approach, recently followed in~\cite{Karlsson18,bana18-10a} is to use instead outage capacity as performance metric.
The outage capacity is an asymptotic performance metric that pertains to the setup in which the channel stays constant (or varies only a finite number of times), as the blocklength grows large---a setting often referred to as \textit{quasi-static} fading.
The use of the outage capacity in~\cite{Karlsson18,bana18-10a} is motivated by the \textit{zero-dispersion} result obtained in~\cite{yang14-07c} which we shall briefly review next.
Let the maximum coding rate be the largest rate at which one can transmit information for a given constraint on the blocklength and the packet error probability.
In~\cite{yang14-07c}, it is shown that the speed at which the maximum coding rate converges to the outage capacity for quasi-static fading channels, as the blocklength increases, is much faster than the speed at which the maximum coding rate converges to Shannon's capacity for nonfading \gls{awgn} channels.
Intuitively, the reason is that errors in quasi-static fading channels are caused by deep-fade events, which cannot be alleviated through coding.

However, this result relies on a Taylor expansion of the maximum coding rate, in which high-order terms that depend on the fading distribution are ignored.
In particular, it is known that these high-order terms become increasingly large as the fading distribution becomes more concentrated around its mean~\cite{yang13-07a}, which is exactly what happens in massive MIMO links when channel hardening occurs.
This makes the use of outage capacity questionable.

Another unsatisfactory consequence resulting from using outage capacity is that the channel can be estimated perfectly at no rate penalty, both in the \gls{ul} and in the \gls{dl}.
Indeed, it is sufficient to transmit a number of pilot symbols that grows sub-linearly with the blocklength~\cite[p.~2632]{biglieri98-10a}.
This is dissatisfying  as the performance of massive MIMO systems in the \gls{urllc} regime are expected to depend heavily on the channel estimation accuracy~\cite{Popovski19}.
This issue is partially addressed in~\cite{Karlsson18,bana18-10a} by utilizing outage-probability approximations in which the rate is multiplied by a correction factor that accounts for pilot overhead.
However, the validity of such approximations is unclear.

\paragraph*{Contributions}
We provide a finite-blocklength framework to analyze the performance of massive MIMO systems in the \gls{urllc} regime.
Specifically, we present finite-blocklength bounds on the error probability that capture the main features of massive \gls{mimo} links, i.e., \gls{ul} pilot transmissions, linear combining/precoding, and mismatched nearest-neighbor detection.
The bounds are based on random coding, pertain to Gaussian codebooks, and rely on the \gls{rcus}~\cite{martinez11-02a}.
Furthermore, by generalizing the analysis presented in~\cite{durisi16-02a,ferrante18-03}, we also obtain  finite-blocklength bounds for the setup in which an (inner) \gls{ostbc} is used at the transmitter side to provide spatial diversity for the case in which \gls{csi} at the transmitter is not available.

We then apply these bounds to two scenarios that are relevant for URLLC:
a \gls{ue}-initiated bidirectional communication link, and a \gls{bs}-initiated bidirectional communication link.
While the first scenario is somewhat standard in massive MIMO analyses, the second scenario is less investigated in the literature.
In the second scenario, similar to the initial-access problem considered in~\cite{Karlsson18}, the \gls{bs} cannot perform beamforming based on \gls{ul}-pilot channel estimation.
Hence, it needs to resort to space-time block-codes to achieve spatial diversity.
Furthermore, as already pointed out in~\cite{Karlsson18}, the significant overhead caused by \gls{dl} pilot transmission prevents the BS from using all available transmit antennas.
For both scenarios, our bounds allow one to determine the optimal number of pilot symbols to be transmitted in order to minimize the SNR required to sustain a target error probability.

\paragraph*{Notation}
Boldface lower-case letters denote vectors and boldface upper-case letters are used for matrices.
We denote by $\veczero_n$ and $\matI_n$, the all-zero vector of size $n$ and the identity matrix of size $n\times n$, respectively.
The superscripts $\tp{(\cdot)}$, $\herm{(\cdot)}$, and $(\cdot)^*$ are used for transposition, Hermitian transposition, and complex conjugation.
The distribution of a standard circularly symmetric Gaussian random variable is denoted by $\cgauss{0}{1}$.
Finally, $\Re(\cdot)$ and $\Im(\cdot)$ denote the real and imaginary part, the expectation operator is denoted by \Ex{}{\cdot}, and the $\ell_2$-norm is written as $\vecnorm{\cdot}$.

\section{Finite-Blocklength Bounds for a Simplified Channel Model}\label{sec:simplified_channel_model}
We start by presenting our finite-blocklength framework for a simplified channel model that, as we shall see, captures the main features of the massive \gls{mimo} setup we are interested in.
Consider the complex-valued additive channel
\begin{equation}\label{eq:simplified_channel}
  v_k=g t_k+ w_k, \quad k=1,\dots,n.
\end{equation}
Here, $t_k$ denotes the channel input, $g$ is a deterministic channel gain, and $w_k$ denotes the additive noise.
The channel output is represented by $v_k$ and $n$ stands for the blocklength.
To transmit the message $m \in \{1,\dots,M\}$, the encoder maps it to one out of $M$ $n$-dimensional codewords $\{\vect(m)\}_{m=1}^{M}$, where $\vect=[t_1,\dots,t_n]$.

The key step to obtain finite-blocklength bounds that are relevant for the massive MIMO setup we are interested in, is to  model appropriately the operations that the decoder is allowed to perform.
In what follows, we will assume that:
\begin{itemize}
  \item The receiver has an estimate $\hat{g}$ of the channel gain $g$ that is treated as perfect.
  \item To decode the transmitted message, the receiver seeks the codeword $\vect(m)$ that, once scaled by $
  \hat{g}$ is the closest to the received vector $\vecv=[v_1,\dots,v_n]$ in Euclidean distance.
  Mathematically, the estimated message $\hat{m}$ at the receiver  is given by
  \begin{equation}\label{eq:mismatched_snn_decoder}
    \hat{m}=\argmin_{\tilde{m}\in \lbrace 1,\dots,M \rbrace} \vecnorm{\vecv-\hat{g}\vect(\tilde m)}^2.
  \end{equation}
\end{itemize}
Some comments are in order.
The receiver just described is the \gls{ml} receiver if and only if $\hat{g}=g$ and $w_k$ is an \iid $\jpg(0,\sigma^2)$ sequence.
This means, that the receiver just introduced treats the additive noise (which is not necessarily Gaussian) as Gaussian.
We refer to this decoder as a mismatched \gls{snn} decoder~\cite{lapidoth02-05a}.

We are now interested in determining a bound on the message error probability $\epsilon$ achieved by this receiver.
To do so, we follow a standard practice in information theory and use a random-coding approach, where we analyze the error probability of an ensemble of random codes, which are generated by drawing the elements of each codeword independently from a given distribution.
Specifically, we consider a Gaussian random code ensemble, where the elements of each codeword are drawn independently from a $\jpg(0,\rho)$ distribution.\footnote{Note that this ensemble is not optimal at finite blocklength, not even when $\hat{g}=g$ and the additive noise is Gaussian~\cite{scarlett16-10a}.
We chose it because it results in simple expressions.
The analysis can be easily extended to other ensembles.}
Here, $\rho$ can be thought of as the average transmit power.
A simple generalization of the random coding union bound in~\cite[Thm.~16]{polyanskiy10-05a} to the mismatched \gls{snn} decoder~\eqref{eq:mismatched_snn_decoder} results in the following bound
\begin{equation}\label{eq:rcu}
  \epsilon\leq \Ex{}{\min\lefto\{1,(M-1)f(\vect,\vecv)\right\}}
\end{equation}
where $f(\vect,\vecv)=\Pr\{\vecnorm{\vecv-\hat{g}\bar{\vect}}^2\leq \vecnorm{\vecv-\hat{g}{\vect}}^2\given \vect,\vecv\}$.
The random variables involved in the bound have the following joint distribution: $P_{\vect,\vecv,\bar{\vect}}(\veca,\vecb,\vecc)=P_{\vect}(\veca)P_{\vecv\given \vect}(\vecb\given \veca)P_{\vect}(\vecc)$.
Coarsely speaking, $\vect$ denotes the transmitted codeword, whereas $\bar{\vect}$ denotes another codeword.
Clearly if $\bar{\vect}$ is closer to $\vecv$ in Euclidean distance after being scaled by $\hat{g}$, the decoded message will be wrong.
The bound~\eqref{eq:rcu} then follows from a tightened version of the union bound.

Although tight, this bound is difficult to compute numerically.
This is because $M$ is typically very large (e.g., $M=2^{50}$ for a code of rate $1/2$ and blocklength $100$).
Hence, the probability term inside the expectation needs to be computed with very high precision---something that is not possible using plain vanilla Monte-Carlo methods.
The approach proposed in~\cite{martinez11-02a} to solve this issue is to upper-bound the probability term using the Chernoff bound.
This results in the so-called \gls{rcus} bound:
\begin{equation}\label{eq:rcus}
  \epsilon\leq \inf_{s>0}\Ex{}{\exp\left(-\max\lefto\{0,\imath_s(\vect,\vecv)-\ln(M-1)\right\} \right) }
\end{equation}
where
\begin{equation}
  \imath_s(\vect,\vecv) = s\vecnorm{\vecv-\hat{g}\vect}^2 -s\frac{\vecnorm{\vecv}^2}{1+s\snr\abs{\hat{g}}^2}-n\ln(1+s\rho\abs{\hat{g}}^2).
\end{equation}
 Next, we will discuss how to use~\eqref{eq:rcus}  to assess the finite-blocklength performance of massive \gls{mimo} links.

\section{\gls{ue}-Initiated Communication}\label{sec:ue_init}
\subsection{Uplink}\label{sec:ue_init_ul}
We assume that transmissions are scheduled using \gls{tdd}.
Each \gls{tdd} frame is divided into an \gls{ul} and a \gls{dl} phase, with each phase lasting for $n$ channel uses.
We assume that $U$ single-antenna \glspl{ue} are simultaneously active and that the \gls{bs} has $B$ antennas.
The \glspl{ue} initiate the transmission by sending orthogonal pilot sequences consisting of $U\leq \np<n$ symbols, each of power $\rul$.
Once the training phase is over, the \glspl{ue} transmit coded data on the remaining $n-\np$ channel uses.

The received signal corresponding to the $k$th transmitted data symbols from the $U$ \glspl{ue} is
\begin{IEEEeqnarray}{rCl}
  \vecy_k &=& \matH \vecx_k + \vecz_k, \quad k=1,\dots, n-
  \np.
\end{IEEEeqnarray}
Here, $\vecx_k\distas \cgauss{\veczero_U}{\rul\matI_U}$ denotes the transmitted symbols from all \glspl{ue} at time $k$, $\matH\in \complexset^{B\times U}$ is the fading matrix, which is random but remains constant over the \gls{tdd} frame, and $\vecz_k\sim \cgauss{\veczero_B}{\matI_B}$ is the \gls{awgn} at the \gls{bs}.
For sake of simplicity, we assume that the entries of $\matH$ are drawn independently from a $\cgauss{0}{1}$ distribution.
However, our framework is general, and can be readily applied to arbitrary fading distributions.

The \gls{bs} uses the $\np$ pilot symbols to estimate the channel matrix.
Throughout the paper, we focus on \gls{mmse} channel estimation, which results in the estimate
\begin{IEEEeqnarray}{rCl}\label{eq:mmse}
  \widehat{\matH} &=& \frac{\sqrt{\np \rul}}{1+\np \rul} \lro{\sqrt{\np \rul}\, \matH + \matZ}.
\end{IEEEeqnarray}
Here, $\matZ$ is a $B\times U$ matrix with \iid $\jpg(0,1)$ entries, which captures the impact of the additive noise on the channel estimate.

Next, the \gls{bs} uses $\widehat{\matH}$ to construct a $B\times U$ linear combiner (e.g., a maximum-ratio combiner) $\matW$ that is used to separate the signals from the $U$ users.
Specifically, the output of the combiner corresponding to the signal transmitted by \gls{ue} $u$ at time $k$ is
\begin{equation}\label{eq:uplink_output_after_combining}
  r_k^{(u)}= \herm{\vecw_u}\vech_u x_k^{(u)}+\sum_{u'\neq u}\herm{\vecw_{u}}\vech_{u'} x_k^{(u')}+\herm{\vecw_u} \vecz_k.
\end{equation}
Here, $\vecw_u$ and $\vech_u$ denote the $u$th column of the matrix $\matW$ and $\matH$, respectively.
Furthermore, $x_k^{(u)}$ stands for the $u$th entry of the vector $\vecx_k$.
Note that the first term in~\eqref{eq:uplink_output_after_combining} corresponds to the desired signal from \gls{ue} $u$, the second term is the residual multiuser interference after linear combining, and the third term is due to additive noise.
Furthermore, note that~\eqref{eq:uplink_output_after_combining} is structurally similar to~\eqref{eq:simplified_channel}:
just set $v_k=r_k^{(u)}$, $t_k=x_k^{(u)}$, $g=\herm{\vecw_u}\vech_u$, and $w_k=\sum_{u'\neq u}\herm{\vecw_{u}}\vech_{u'} x_k^{(u')}+\herm{\vecw_u} \vecz_k$.

We assume that the \gls{bs} decodes the message from each \gls{ue} separately (no joint decoding).
Furthermore, we assume that the \gls{bs} treats the acquired channel estimate as perfect, and the residual multiuser interference as additive noise.
In the notation introduced in Section~\ref{sec:simplified_channel_model}, this corresponds to performing mismatched \gls{snn} decoding with $\hat{g}=\herm{\vecw}_u\widehat{\vech}_u$.
It follows that the error probability bound~\eqref{eq:rcus} applies to this setup, once the substitutions described above are performed, and after taking an additional expectation over $\matH$ and over  $\matZ$ in~\eqref{eq:mmse}.
Indeed, different from the setup in Section~\ref{sec:simplified_channel_model}, the channel is now random.

\subsection{Downlink}
In the \gls{dl} phase, the \gls{bs} multiplies the $U$-dimensional symbol vector $\vecx_k\sim \cgauss{\veczero_U}{\rdl \matI_U}$ at time $k$   by the $B \times U$ linear precoding matrix $\matP$, constructed on the basis of the channel estimate $\widehat{\matH}$ obtained in the \gls{ul} phase.
We assume that each column of $\matP$ is normalized so that the expected value of its $\ell_2$ norm is $1$.
The received signal at \gls{ue} $u$ corresponding to the $k$th transmitted data vector from the \gls{bs} is
\begin{IEEEeqnarray}{rCl}\label{eq:downlink_output}
  y_k^{(u)} &=&  \tp{\vech_u} \vecp_u x^{(u)}_k + \sum_{u'\neq u} \tp{\vech_{u}} \vecp_{u'} x^{(u')}_k + z^{(u)}_k.
\end{IEEEeqnarray}
Here, $\vecp_u$ denotes the $u$th column of the linear precoding matrix $\matP$ and $z^{(u)}_k\sim \cgauss{0}{1}$ denotes the
\gls{awgn} at \gls{ue} $u$.
Similar to~\eqref{eq:uplink_output_after_combining}, the first term in~\eqref{eq:downlink_output} corresponds to the desired signal from the \gls{bs} while the second term contains the residual multiuser interference after linear precoding.
Again, we can put~\eqref{eq:downlink_output} in the form given in~\eqref{eq:simplified_channel} by setting $v_k = y_k^{(u)}$, $t_k=x^{(u)}_k$, $g=\tp{\vech_u} \vecp_u$, and $w_k=\sum_{u'\neq u} \tp{\vech_{u}} \vecp_{u'} x^{(u')}_k + z^{(u)}_k$.

We assume that each \gls{ue} performs mismatched \gls{snn} decoding where the multiuser interference is treated as noise.
Since no pilot symbols are transmitted in the downlink phase, no knowledge of the channel gain $g=\tp{\vech_u} \vecp_u$ is available at the \glspl{ue}.
We assume, however, that each \gls{ue} has some statistical knowledge of the channel; specifically, as commonly done in the massive \gls{mimo} literature, we assume that each \gls{ue} knows the mean $\Ex{}{\tp{\vech}_u \vecp_u}$ of the channel gain and uses this quantity to perform mismatched \gls{snn} decoding.
Specifically, we set $\hat{g}=\Ex{}{\tp{\vech}_u \vecp_u}$.
Obvioulsy, channel hardening is critical for this choice to result in good performance.

The error probability in the \gls{dl} can be readily evaluated using~\eqref{eq:rcus} after taking an additional expectation over $\matH$ and over the matrix $\matZ$ in~\eqref{eq:mmse}.

\section{\gls{bs}-Initiated Communication}

In this section, we consider a \gls{bs}-initiated bidirectional communication link.
In the \gls{dl}, the \gls{bs} needs to deliver a common message to all \glspl{ue}.
Each \gls{ue} then replies individually with a potentially distinct message.
We will focus in this section exclusively on the first \gls{dl} phase, since the \gls{ul} phase is similar to the one described in Section~\ref{sec:ue_init_ul}.

The initial \gls{dl} phase is challenging, since no \gls{csi} is available to the \gls{bs}.
Hence, no beamforming is possible.
This means that the spatial diversity required to achieve the target reliability needs to be provided through the use of space-time codes.
Furthermore, the \glspl{ue} cannot rely on channel hardening in the decoding process, and instead need to estimate (implicitly or explicitly) the fading channel.
In this section, we consider explicit channel estimation based on downlink pilot symbols.

As noted previously in the massive \gls{mimo} literature (see~\cite{Karlsson18} and references therein), it is not feasible to transmit orthogonal pilot sequences from all available antennas.
Indeed, in short-packet transmissions, the blocklength $n$ may be of the same order as the number of available antennas $B$, which makes orthogonal pilot transmission from all antennas unattractive since too few resources would be left for the transmission of the data symbols.

Following the strategy in~\cite{Meng16,Karlsson18}, we assume that the \gls{bs} relies on an \gls{ostbc} that uses only $B'$ of the $B$ available \gls{bs} antennas.
Our aim is to use the error probability bound~\eqref{eq:rcus} in order to characterize the trade-off between $B'$ and $\np$.
On the one hand, increasing $B'$ results in more spatial diversity, which lowers the error probability; on the other hand, increasing $B'$ results in an increased pilot overhead, which yields to a reduction of the number of symbols that can be used for data transmission.

To adapt~\eqref{eq:rcus} to the scenario just described, we will use the \gls{ostbc} to space-time encode the coded symbols generated by a Gaussian random code.
Then, we will apply the error probability bound in~\eqref{eq:rcus} to characterize the performance achievable using downlink pilot transmission and mismatched \gls{snn} decoding.

We assume that the \gls{ostbc} produces matrix-valued symbols ${\matX} \in \complexset^{ B'\times n\sub{c}}$, each one encoding $n\sub{s}\leq\nc$ complex-valued input symbols $\lbrace q_i \rbrace_{i=1}^{n\sub{s}}$ generated independently from a $\cgauss{0}{\rdl/B'}$ distribution.
Each \gls{ostbc} codeword is transmitted over $n\sub{c}$ channel uses and across $B'$ antennas where ${B' \ll B}$.
The rate of the \gls{ostbc} is given by $R\sub{ostbc}=n\sub{s}/\nc$.
We follow~\cite{Larsson08} and express each \gls{ostbc} symbol ${\matX}$ as
\begin{IEEEeqnarray}{rCl}
  {\matX} &=& \sum_{i=1}^{n\sub{s}} \Re(q_i) \matA_i + j\Im(q_i) \matB_i.
\end{IEEEeqnarray}
The orthogonality assumption implies that
\begin{IEEEeqnarray}{rCl}
  \Ex{}{\matX\herm{{\matX}}} &=& \frac{\nc \rdl}{B'} \matI_{B'}.
\end{IEEEeqnarray}
Hence, $\rdl$ can be thought of as the total transmit power in each time instant.
Similar to~\cite{Karlsson18}, for $B'=4$, we choose $\{\matA_i,\matB_i\}$ so that the resulting \gls{ostbc} is the one given in~\cite[Example 7.4]{Larsson08} and for larger values of $B'$ we construct $\{\matA_i,\matB_i\}$ following the procedure outlined in~\cite{tarokh99-07a} (although a higher-rate \gls{ostbc} might be available).

A dimension-reducing matrix $\matU$ of size $B \times B'$ is used to map each \gls{ostbc} symbol ${\matX}$ to the $B$ \gls{bs} antennas.
For simplicity, we assume that this matrix is obtained by eliminating the last $B-B'$ columns of a randomly generated unitary matrix of dimension $B \times B$.\footnote{Better designs may be possible, especially if information about the statistical properties of the propagation channel is available at the \gls{bs}.}

The \gls{dl} transmission consists of a training phase and a data phase.
In the training phase, orthogonal pilot sequences of length $n\sub{p}\geq B'$ are transmitted from each \gls{bs} antenna.
These pilot sequences are used at each \gls{ue} to estimate the effective channel $\vech_u^{(\text{eff})}=\tp{\matU}\vech_u$, where $\vech_u \in \complexset^{B}$ denotes the channel from the $B$ antennas at the \gls{bs} to
\gls{ue} $u$, and $\vech_u^{(\text{eff})} \in \complexset^{B'}$.
We assume that channel estimation is performed using the \gls{mmse} principle (see~\eqref{eq:mmse}), and denote by
$\widehat{\vech}_u^{(\text{eff})} \in \complexset^{B'}$ the \gls{mmse} estimate.

In the data phase, $\ell$ space-time-coded symbols are transmitted from the \gls{bs}.
We assume that the overall \gls{dl} phase lasts at most $n$ channel uses.
Hence, for a given choice of the \gls{ostbc}, the integers $n\sub{p}$ and $\ell$ need to be chosen such that $\np + \ell\nc \leq n$.
The received signal at the $u$th \gls{ue} corresponding to the $k$th \gls{ostbc} symbol ${\matX}_k$, $k=1,\dots,\ell$, is given by
\begin{IEEEeqnarray}{rCl}\label{eq:ostbc_rx}
  \vecy^{(u)}_k &=& \tp{\vech_u}\matU \matX_k  + \vecz_k^{(u)} = \tp{(\vech^{\text{eff}}_u)} {\matX}_k+ \vecz_k^{(u)}.
\end{IEEEeqnarray}
Here, $\vecy^{(u)}_k$ is an $\nc$-dimensional vector, and the additive noise is denoted by $\vecz_k^{(u)}\distas \jpg(\veczero_{\nc}, \matI_{\nc})$.

We assume that the $u$th \gls{ue} obtains an estimate $r^{(u)}_{k,i}$ of the $i$th coded symbol $q_{k,i}$ transmitted on the $k$th \gls{ostbc} symbol as follows~\cite{Larsson08}:
\begin{IEEEeqnarray}{rCl}\label{eq:ostbc_stat}
  r^{(u)}_{k,i} &=& \Re\lro{ \frac{ \tp{(\widehat{\vech}^{\text{eff}}_u)}\matA_i }{\vecnorm{\widehat{\vech}^{\text{eff}}_u}} \herm{(\vecy^{(u)}_k)}} \notag \\ &&+ j \Im\lro{ \frac{ \tp{(\widehat{\vech}^{\text{eff}}_u)}\matB_i }{\vecnorm{\widehat{\vech}^{\text{eff}}_u}} \herm{(\vecy^{(u)}_k)}}.
\end{IEEEeqnarray}
Then mismatched \gls{snn} decoding on the basis of $\widehat{\vech}^{\text{eff}}_u$ is performed with $\hat{g}=\vecnorm{\widehat{\vech}_u^{\text{eff}}}$, channel inputs given by the transmitted symbols $\{q_{k,i}\}$, and channel outputs given by the corresponding estimates $\{r^{(u)}_{k,i}\}$.
We refer the reader to~\cite{Karlsson18} for a decomposition of~\eqref{eq:ostbc_stat} into useful-signal part and  intersymbol-interference terms that result from channel-estimation errors.

\section{Numerical Results}

In this section, we consider a scenario where $B=100$ and $U=10$.
Furthermore, we assume that $n=288$ and that $\log_2M = 30$, i.e., each message consists of $30$ bits.
These two values are in agreement with the so-called compact downlink control information transmission scenario~\cite{R1-1720997-Ericsson}.
We assume, throughout this section, an \iid Rayleigh fading channel.

First, we consider the \gls{ue}-initiated-transmission scenario and assume a target error probability of  $\epsilon=10^{-5}$ on the bi-directional link.
To satisfy the error-probability target, we require $\epsilon\sub{ul}=\epsilon\sub{dl}=\epsilon/2$, where $\epsilon\sub{ul}$ and $\epsilon\sub{dl}$ denote the error probability on the \gls{ul} and \gls{dl}, respectively.

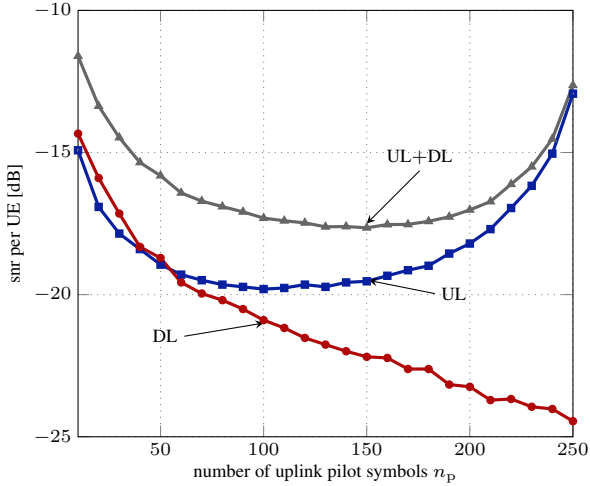
\begin{figure}[t]
    \centering
%
%
%
%
%
%
%

\begin{tikzpicture}
  \pgfplotsset{
      scaled y ticks = false,
      width=\fwidth*0.45,
      height=\fwidth*0.4,
       title style={yshift=-6pt,}
  }
  \begin{axis}[
    xmin=10,
    xmax=250,
    xlabel={number of uplink pilot symbols $\np$},
    ymin=-25,
    ymax=-10,
    ylabel={snr per UE [dB]},
    grid=major
    ]


    \addplot [ color=blue!80!black, mark=square, mark size=1pt, line width = 1.2pt] table [y index={1}, x index = {0}, col sep=comma] {./figures/Pilots_vs_SNR_B_100_U_10_MRC_epsilon_1e-5_ULDL.csv};

    \addplot [ color=red!80!black, mark=*, mark size=1pt, line width = 1.2pt] table [y index={2}, x index = {0}, col sep=comma] {./figures/Pilots_vs_SNR_B_100_U_10_MRC_epsilon_1e-5_ULDL.csv};

    \addplot [ color=gray!80!black, mark=triangle, mark size=1pt, line width = 1.2pt] table [y index={3}, x index = {0}, col sep=comma] {./figures/Pilots_vs_SNR_B_100_U_10_MRC_epsilon_1e-5_ULDL.csv};

    \coordinate (pt1) at (axis cs: 151,-19.5);
    \coordinate (pt2) at ($(pt1) + (25pt,-5pt)$);
    \draw[<-] (pt1)--(pt2) node[anchor = west] at ($(pt2) + (0pt,0pt)$) {UL};

    \coordinate (pt1) at (axis cs: 100,-21);
    \coordinate (pt2) at ($(pt1) + (-30pt,-5pt)$);
    \draw[<-] (pt1)--(pt2) node[anchor=east] at ($(pt2)$) {DL};

    \coordinate (pt1) at (axis cs: 150,-17.5);
    \coordinate (pt2) at ($(pt1) + (20pt,20pt)$);
    \draw[<-] (pt1)--(pt2) node[anchor=south] at ($(pt2) + (0pt,0pt)$) {UL$+$DL};
  \end{axis}

\end{tikzpicture}

%
        \caption{The minimum SNR required in the \gls{ul} and in the \gls{dl} to achieve $\epsilon\sub{ul}=\epsilon/2$ and $\epsilon\sub{dl}=\epsilon/2$, respectively. Here, $\epsilon = 10^{-5}$, $B=100$, $U=10$, $n=288$, and $\log_2M=30$. The propagation channel is modeled as \iid spatially-white Rayleigh fading.}
        \label{fig:ue_init}
\end{figure}

In Fig.~\ref{fig:ue_init}, we illustrate the minimum SNR (obtained via~\eqref{eq:rcus}) required for both the \gls{ul} and the \gls{dl} transmission to achieve $\epsilon\sub{ul}$ and $\epsilon\sub{dl}$, respectively.
In the figure, we assume maximum-ratio combining and maximum-ratio precoding.
We see that the \gls{ul} SNR decreases as the number of pilot symbols increase up until $\np= 100$.
For $\np>100$, the required SNR increases because the channel estimation overhead offsets the performance gain resulting from a more accurate channel estimate.
Not surprisingly, the picture is different for the \gls{dl}.
Since pilot overhead penalizes only the \gls{ul}, the downlink SNR decreases as the number of pilot symbols increase.
We also see from the figure that the optimum number of pilot symbols that minimizes the total SNR is $n\sub{p}=150$.

\begin{figure}[t]
    \centering
%
%
%
%
%
%
%

\begin{tikzpicture}
  \pgfplotsset{
      scaled y ticks = false,
      width=\fwidth*0.45,
      height=\fwidth*0.4,
      title style={yshift=-6pt,}
  }
  \begin{axis}[
    xmin=12,
    xmax=250,
    xlabel={number of pilot symbols},
    ymin=0,
    ymax=13,
    ylabel={snr [dB]},
    grid=major,
    ]


    %

    \addplot+ [ mark=*, mark size=1pt, line width = 1.2pt] table [y index={1}, x index = {0}, col sep=comma] {./figures/Pilots_vs_SNR_B_4_U_10_k_30_MRC_epsilon_1e-05_Saddle_ns_3_OSTBC.csv};

    \coordinate (pt3) at (axis cs: 24,9.1);
    \coordinate (pt4) at ($(pt3) + (15pt,15pt)$);
    \draw[<-] (pt3)--(pt4) node[text width=4cm] at ($(pt4) + (40pt,5pt)$) {$(B',n\sub{s},R\sub{ostbc})=(4,3,3/4)$};

    \addplot+ [  mark=*, mark size=1pt, line width = 1.2pt] table [y index={1}, x index = {0}, col sep=comma] {./figures/Pilots_vs_SNR_B_8_U_10_k_30_MRC_epsilon_1e-05_Saddle_ns_8_OSTBC.csv};

    \coordinate (pt3) at (axis cs: 32,3.87);
    \coordinate (pt4) at ($(pt3) + (15pt,5pt)$);
    \draw[<-] (pt3)--(pt4) node[text width=2cm] at ($(pt4) + (30pt,3pt)$) {$(8,8,1/2)$};

    \addplot [only marks, color=blue, mark=square*, mark size=1pt, line width = 1.2pt] table [y index={1}, x index = {0}, col sep=comma] {./figures/Pilots_vs_SNR_B_10_U_10_k_30_MRC_epsilon_1e-05_Saddle_ns_32_OSTBC.csv};

    \coordinate (pt3) at (axis cs: 96,1.91);
    \coordinate (pt4) at ($(pt3) + (10pt,-5pt)$);
    \draw[<-] (pt3)--(pt4) node[anchor=west,text width=2.5cm] at ($(pt4) + (0pt,0pt)$) {$(10,32,1/2)$};

    \addplot [only marks, color=green, mark=*, mark size=1pt, line width = 1.2pt] table [y index={1}, x index = {0}, col sep=comma] {./figures/Pilots_vs_SNR_B_12_U_10_k_30_MRC_epsilon_1e-05_Saddle_ns_64_OSTBC.csv};
    \coordinate (pt3) at (axis cs: 160,2.6);
    \coordinate (pt4) at ($(pt3) + (20pt,10pt)$);
    \draw[<-] (pt3)--(pt4) node[text width=2.5cm] at ($(pt4) + (35pt,3pt)$) {$(12,64,1/2)$};

  \end{axis}

\end{tikzpicture}

%
        \caption{The minimum SNR required for $B=100$ and $n=288$ to achieve $\epsilon\sub{ostbc}=10^{-5}$ as a function of the number of pilot symbols $\np$ per antenna, for four different \glspl{ostbc}. The propagation channel is modeled as \iid spatially-white Rayleigh fading.}
        \label{fig:bs_init}
\end{figure}
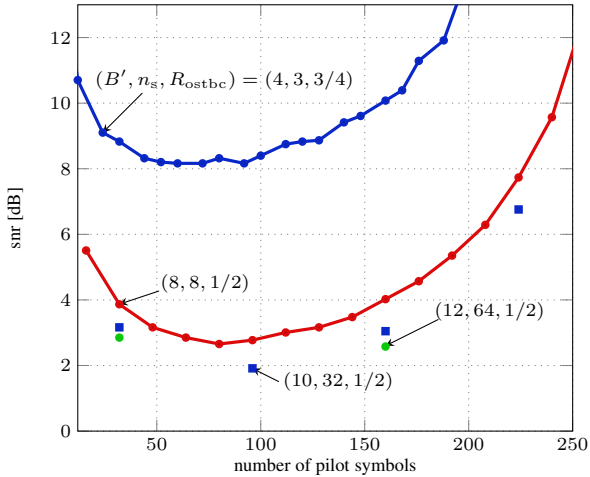

Next, we consider the \gls{bs}-initiated-transmission scenario.
In Fig.~\ref{fig:bs_init},  we illustrate the minimum SNR (obtained via~\eqref{eq:rcus}) required to achieve $\epsilon=10^{-5}$, for four different \glspl{ostbc} as a function of the number of pilot symbols transmitted from each active \gls{bs} antenna.
It can be seen that setting $B'=4$ results in a high required SNR, because this space-time code offers very limited spatial diversity.
As we increase $B'$ to $10$, the required minimum SNR can be reduced by about $6.3\dB$.
Increasing $B'$ further is not helpful because of the channel-estimation overhead.
In particular, we see from the figure that, as the number of active antennas increases, setting $\np$ appropriately is critical.
For example, for $B'=10$, the required SNR to achieve $10^{-5}$ is about $1.9\dB$ when $\np=96$ but $6.7\dB$ when $\np=224$.


\section{Conclusion}
We have presented a framework based on finite-blocklength information theory that is suitable for determining the error probability achievable on massive \gls{mimo} links in \gls{urllc} scenarios.
Through numerical simulations involving bi-directional \gls{ue}-initiated and \gls{bs}-initiated communication links, we have illustrated how to use this framework  to optimize the number of pilot symbols to minimize the transmit power given a reliability and a latency constraint.

 \bibliographystyle{IEEEtran}
 \bibliography{./Inputs/giubib}
\end{document}